\documentclass[12pt]{article}
\usepackage{a4,epsfig,axodraw}

\newcommand{\newc}{\newcommand}
\newc{\unit}[1]{\,\mathrm{#1}}
\newc{\op}[1]{\mathrm{#1}}
\newc{\beq}{\begin{equation}}
\newc{\eeq}{\end{equation}}
\newc{\barr}{\begin{eqnarray}}
\newc{\earr}{\end{eqnarray}}
\newc{\eps}{\epsilon}
\newc{\const}[1]{{#1}}
\newc{\prtcle}[1]{\mathrm{#1}}
\newc{\var}[1]{{#1}}
\newc{\mtext}[1]{\mathrm{#1}}
\newc{\ttimes}[1]{\times 10^{#1}}
\newc{\name}[1]{\mbox{#1}}
\newc{\mycap}[3]{\caption[#3]{{\it\label{#1}{#2}}}}
\newc{\mycaps}[2]{\caption[#2]{{\footnotesize\label{#1}{#2}}}}
\newc{\dgr}[1]{{#1}^{\circ}}
\newc{\rpv}{{\not \!\! R_p}}
\newc{\gsim}{\stackrel{>}{\sim}}
\newc{\lsim}{\stackrel{<}{\sim}}
\newc{\ra}{\rightarrow}
\newc{\tp}{t^\prime}
\newc{\tpp}{t^{\prime\prime}}
\newc{\spr}{s^\prime}
\newc{\spp}{s^{\prime\prime}}
\newc{\up}{u^\prime}
\newc{\upp}{u^{\prime\prime}}
\newc{\tps}{t^{\prime2}}
\newc{\ups}{u^{\prime2}}
\newc{\lam}{\lambda}
\def\xi{\ensuremath{\chi^{+}}}

\def\slep{\ensuremath{\widetilde{l}}}

\def\snu{\ensuremath{\widetilde{\nu}}}

\def\lle{\ensuremath{\rm LL\bar{E}}}

\def\gev{\ensuremath{\unit{GeV}}}

\def\gevcc{\unit{GeV/c^2}}

\def\pbi{\unit{pb^{-1}}}
\def\fb{\unit{fb}}
\def\mrad{\unit{mrad}}
\def\gg{\ensuremath{\gamma \gamma}}
\def\ra{\ensuremath{\rightarrow}}
\def\epem{\ensuremath{e^+e^-}}

\begin{document}
\begin{center}
\vspace{2cm}
\title{Single Sneutrino/Slepton Production
  at LEP2 \\ and the NLC}
\author{B.C. Allanach$^1$, H. Dreiner$^1$, P. Morawitz$^2$ and M.D. 
Williams$^2$}
\date{{\small $^1$ Rutherford Lab., Chilton, Didcot, OX11 0QX, UK\\
$^2$ Imperial College, HEP Group, London SW7 2BZ, UK}}
\maketitle

\end{center}

\vspace{3ex}
\begin{abstract}
We propose a new method of detecting supersymmetry at LEP2 and the NLC when
R-parity is violated by an $\lle$ operator. We consider the processes
$e\gamma\ra\snu_j e_k$ and $e\gamma \ra \slep_k \bar{\nu}_j$ which can test
seven of the nine $\lle$-operators. A Monte-Carlo analysis is performed to
investigate the sensitivity to the sneutrino signal, and the 5$\sigma$
discovery contours in the $m_{\tilde{\nu}_j}$ vs. $\lam$ plane are presented.
For an integrated luminosity of $100\pbi$, sneutrinos with masses up to
$M_{\tilde \nu}<170\gevcc$ could be discovered in the near future at LEP2. For
the charged slepton production the cross-section is too low to be detectable. 
\end{abstract}

\section{Introduction}
In the R-parity violating version of the minimal supersymmetric standard model
\cite{review} supersymmetric particles can be produced singly. The production 
cross section is suppressed by the $\rpv$ Yukawa coupling but the kinematic
reach is typically twice that of supersymmetric pair production mechanism. 
Here, we consider the possible detection of R-parity violating
supersymmetry at LEP2 through the superpotential terms\footnote{Here $L$ and $E$
are the SU(2) doublet and singlet lepton superfields respectively.} 
\beq 
W_{LLE} =\frac{1}{2} \lam_{ijk} (\eps_{\alpha\beta}L_i^\alpha L_j^\beta) E^c_k, 
\label{LLEsup}
\eeq 
where $i,j,k=1,2,3$ are family indices and $\alpha,\beta=1,2$ are SU(2)-gauge
indices. $\eps_{\alpha\beta}$ is the totally antisymmetric tensor, $\eps_{12}
=+1$.

The single {\it resonant} production of sneutrinos at $e^+e^- $-colliders was
first considered in \cite{hall,bgh}. More detailed studies for specific
colliders (LEP2, NLC) were performed in \cite{resonance} and experimental analyses
have since been performed \cite{experiment}. All of these studies
were based on the assumption of a single dominant coupling. If this assumption
is dropped the resonantly produced sneutrino can also decay via other
non-vanishing operators \cite{erler,kalinowski}.  All of these previous studies
were restricted to the operators $L_1L_{2,3}{\bar E}_1$. In \cite{kalinowski}
the indirect effects of the t-channel exchange of a sneutrino were studied for
dominant $L_1L_{2,3}{\bar E}_1$, as well. This could easily be extended to other
operators $LLE^c$. However, the effect is too small to be relevant. 

None of the previous direct searches considered the operators
\beq
L_1 L_2 E^c_2, \; L_1 L_2 E^c_3, \; L_1 L_3 E^c_2,\; L_1 L_3 E^c_3,\;
L_2 L_3 E^c_1, \label{newops}
\eeq
as well as $L_2L_3{\bar E}_{2,3}$. In this letter we investigate the possibility
of directly testing the operators (\ref{newops}) via sneutrino or slepton 
production 
\barr
\gamma(p_1)+e^\pm(p_2)&\ra& e^\pm_k(q_1) +{\tilde\nu}_j(q_2),\label{snu}\\ 
\gamma(p_1)+e^\pm(p_2)&\ra&  {\tilde e}_j^\pm(q_1)+\nu_k(q_2).\label{sel} 
\earr
In parentheses we have included the 4-momenta used in the calculation below. 
These processes are competitive for $e_k=\mu,\,\tau$ and ${\tilde\nu}_j={\tilde
\nu}_\mu,\,{\tilde\nu}_\tau$. Due to the incoming photon we are only restricted 
to one electron index. The final state charged lepton/slepton can be an SU(2)
doublet or an SU(2) singlet. The indices $j,k=1,2,3$ are free. Thus we can in
principle also study the operators $L_1L_{2,3}E^c_1$ but here the resonant 
sneutrino production is more sensitive. For simplicity we shall only consider 
one non-zero $\rpv$ coupling at a time. The operators $LLE^c$ can also be
tested at hadron colliders via various direct and indirect signatures 
\cite{hadron}.

\section{Cross Section Evaluation} 
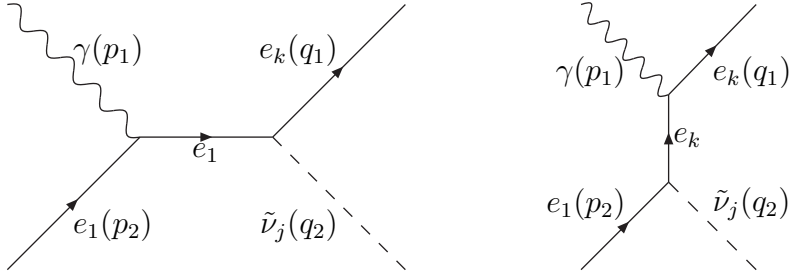
\begin{figure}
\begin{center}
\begin{picture}(300,100)(0,0)
\put(0,0){\begin{picture}(150,100)(0,0)
\Photon(0,100)(50,50){3}{5}
\ArrowLine(0,0)(50,50)
\ArrowLine(50,50)(100,50)
\ArrowLine(100,50)(150,100)
\DashLine(100,50)(150,0){5}
\Text(75,48)[tc]{$e_1$}
\Text(25,23)[tl]{$e_1(p_2)$}
\Text(25,77)[bl]{$\gamma(p_1)$}
\Text(125,23)[tr]{$\tilde{\nu}_j(q_2)$}
\Text(125,77)[br]{$e_k(q_1)$}
\end{picture}}
\put(200,0){\begin{picture}(100,100)(0,0)
\Photon(17,100)(50,66){3}{5}
\ArrowLine(17,0)(50,33)
\ArrowLine(50,33)(50,66)
\ArrowLine(50,66)(83,100)
\DashLine(50,33)(83,0){5}
\Text(52,50)[cl]{$e_k$}
\Text(34,19)[br]{$e_1(p_2)$}
\Text(34,81)[tr]{$\gamma(p_1)$}
\Text(67,19)[bl]{$\tilde{\nu}_j(q_2)$}
\Text(67,81)[tl]{$e_k(q_1)$}
\end{picture}}
\end{picture}
\end{center}
\caption{Feynman diagrams contributing to the $\gamma e\ra e_k{\tilde\nu}$ 
sub-process.}
\label{fig:snuprod}
\end{figure}

The tree-level Feynman diagrams for the sneutrino production (\ref{snu}) are 
shown in Figure~\ref{fig:snuprod}. The spin averaged matrix element squared for 
this process in the Weizs\"acker-Williams approximation \cite{WW} (on-shell 
photon) is given by\footnote{This matrix element squared was first presented
in \cite{proc} in the limit of a massless final state charged lepton $m_{e_k}=0.
$} 
\begin{eqnarray}
|\bar{M}|^2 &=& e^2 \lam^2 \left[\frac{\up(\tp-3 m_{e_k}^2)}{\spr(\tp-m_{e_k}^2
)} + \frac{3m_{e_k}^4-\tps+2\ups}{2\spr(\tp-m_{e_k}^2)} + \right. \nonumber \\
&& \left. \frac{8m_{e_k}^4+ \tp(2\tp+2\up-\spr)-m_{e_k}^2(5\spr+4\tp+8\up)}
{2(\tp-m_{e_k}^2 )^2} \right].
\label{LLEmat}
\end{eqnarray}
The symbol $e$ denotes the absolute value of the charge of the electron and 
$\lam$ is the $\rpv$ Yukawa coupling. $m_{e_k}$ is the mass of the charged
lepton $e_k$ and we have neglected the mass of the incoming electron. We have 
made use of the Mandelstam variables 
\begin{eqnarray}
\spr &=& (p_1+p_2)^2 = 2 p_1\cdot p_2, \nonumber \\
\tp &=& (p_1-q_1)^2 = -2p_1\cdot q_1+m_{e_k}^2, \label{mandel}\\
\up &=& (p_1-q_2)^2 = -2p_1\cdot q_2+m_{\tilde{\nu}}^2, \nonumber 
\end{eqnarray}
where $m_{\tilde{\nu}}$ is the mass of the produced sneutrino and the 
four-momenta are defined in Eq.(\ref{snu}) and Figure~\ref{fig:snuprod}. 

\begin{figure}
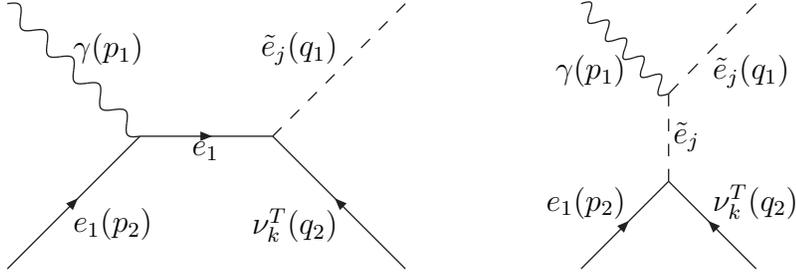

\begin{center}
\begin{picture}(300,100)(0,0)
\put(0,0){\begin{picture}(150,100)(0,0) \input{selchan}
\end{picture}}
\put(200,0){\begin{picture}(100,100)(0,0) \input{telchan}
\end{picture}}
\end{picture}
\end{center}
\caption{Feynman diagrams contributing to the $\gamma e\ra {\tilde e}_j \nu_k$
subprocess. Arrows on fermion lines denote flow of fermion number not momentum.}
\label{fig:selprod}
\end{figure}

The tree-level diagrams for the selectron production (\ref{sel}) are shown in
Figure~\ref{fig:selprod}. The spin-averaged matrix-element squared is given
in the Weizs\"acker-Williams approximation \cite{WW} by
\begin{equation}
|\bar{M}|^2 = \frac{e^2 \lam^2}{2}  \left[
\frac{\tpp(\tpp+m_{\tilde{e}}^2)}{(\tpp-m_{\tilde{e}}^2)^2} - \frac{\upp}{\spp} 
+\frac{\tpp (2 m_{\tilde{e}}^2 - \spp)} {\spp(\tpp-m_{\tilde{e}}^2)} \right],
\label{selprod}
\end{equation}
where $m_{\tilde{e}}$ is the mass of the produced selectron. We have neglected
the electron mass. The Mandelstam variables are as in Eqs.(\ref{mandel}) with
the replacements $(\spr,\tp,\up)\ra(\spp,\tpp,\upp)$ combined with $m_{\tilde
\nu}\ra0,\, m_{e_k}\ra m_{\tilde e}$.

\begin{figure}
\centering 
\epsfig{file=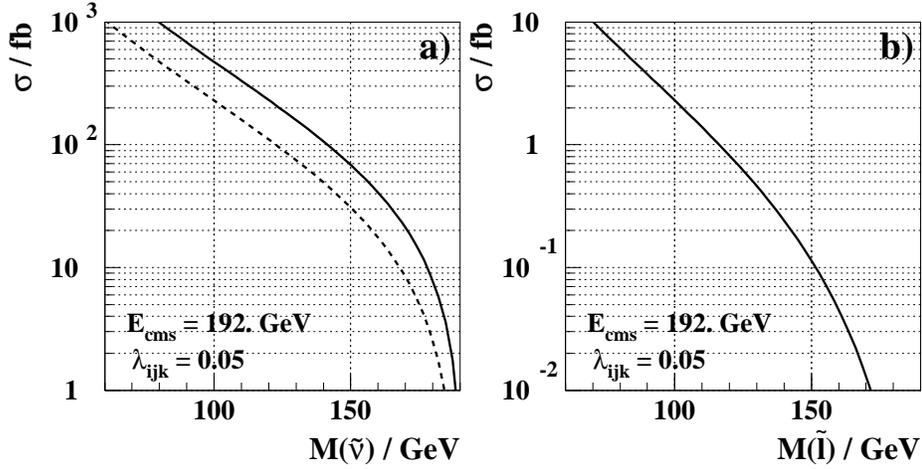,width=14cm} 
\caption{The cross-sections for single production of a) sneutrinos and b) 
charged sleptons at a centre-of-mass energy of $192\gev$ and 
for $\lam_{ijk} = 0.05$ as a function of the sneutrino/slepton mass. In a) 
the solid line is the cross-section for $\epem \ra e\mu\snu$ and 
the dashed line is the cross-section for $\epem \ra e\tau\snu$.}
\label{single.xsect} 
\end{figure}

\begin{figure}
\centering 
\epsfig{file=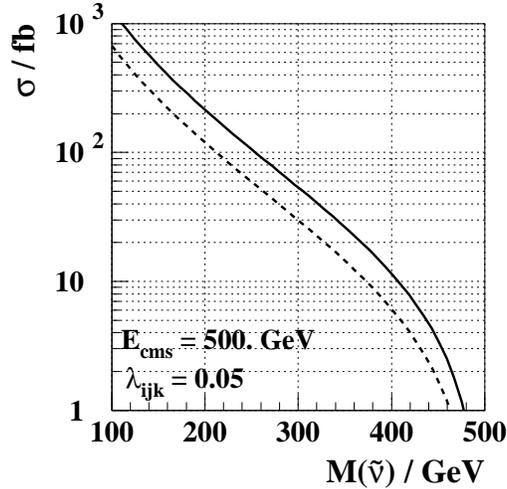,width=7cm} 
\caption{The cross-section for single production of sneutrinos at a Next Linear
Collider with a centre-of-mass energy of $500\gev$ as a function of the 
sneutrino mass. The solid line is the cross-section for $\epem\ra e\mu\snu$ and 
the dashed line is the cross-section for $\epem \ra e\tau\snu$.} 
\label{nlc} 
\end{figure}

The total cross-section for the $e\gamma$ subprocesses may be written as
\begin{equation}
\sigma(s^\prime; \, e\gamma) = 
     \int^{t_{max}}_{t_{min}} \frac{1}{16\pi s^{\prime2}} |\bar{M}|^2 dt,
\label{egxsect}
\end{equation}
where  
\begin{eqnarray}
t_{max/min} & = & \left( \frac{m_{q_2}^2-m_{q_1}^2}{2\sqrt{\spr}} \right)^2 -
\left( \frac{\sqrt{\spr}}{2} \mp \sqrt{ \left( \frac{\spr+m_{q_1}^2-m_{q_2}^2}
{2\sqrt{\spr}} \right)^2 - m_{q_1}^2} \,\right)^2,
\end{eqnarray}
where $m_{q_1}$ and $m_{q_2}$ are the masses of the particles with momenta
$q_1$ and $q_2$, and  the variable $s'$ (or $\spp$ for selectrons)
 is defined in Eqs.(\ref{mandel}). 
 The cross-sections for $\epem \ra el\snu$ and $\epem \ra e\nu
\slep$ are obtained from the $e\gamma$ cross-section by
\begin{equation}
\sigma(s; \, \epem) = 2 \int_0^1 f_\gamma(y) \; \sigma(ys; \, e\gamma) dy,
\end{equation}
where $f_\gamma(y)$ is the photon distribution in the electron at a given 
fraction, $y$, of the electron momentum. The factor of two is due to the charge
conjugate diagrams when an anti-sneutrino or anti-charged slepton is produced. 
We use the following version of the Weizs\"{a}cker-Williams distribution 
\cite{Frixione}: 
\begin{eqnarray}
f_\gamma(y) & = & \frac{\alpha_{em}}{2\pi} \left\{ 2(1-y)
   \left[\frac{m_e^2 y}{E^2(1-y)^2\theta_c^2 + m_e^2y^2} - \frac{1}{y} \right] 
     \right. \nonumber \\
 &   & \left. + \frac{1+(1-y)^2}{y} \log\frac{E^2(1-y)^2\theta_c^2 + m_e^2y^2}
{m_e^2y^2} \right\}, \label{ww.aprox}
\end{eqnarray}
where $\theta_c$ is the maximum scattering angle of the beam electron and $E$
is the beam energy. The value of $\theta_c$ is taken to be $30 \mrad$ -   a
typical value for the coverage of the luminosity monitors in the LEP
experiments - for two reasons: Firstly because for large values of $\theta_c$
the photon emitted from the beam electron is no longer on-shell, and the
validity of the  Weizs\"{a}cker-Williams approximation is compromised. And
secondly because it simplifies the feasability study presented in
Section~\ref{analysis.sec}, which  considers three lepton topologies, assuming
that the beam electron deflected at small angles is not measured in the
detector. If we were to allow for the full range of $\theta_c$ in
Eq.(\ref{ww.aprox}), the cross-section would be larger by approximately $20\%$.
The cross-sections for $\epem \ra e\mu\snu$ and $\epem \ra e\tau\snu$ are shown
as a function of the slepton and sneutrino mass in Figure~\ref{single.xsect}
for $\lam_{ijk} = 0.05$. The cross-section for single  slepton production is
much smaller than that for single sneutrino production, since the t-channel
diagram (which is the dominant diagram) is suppresed. Single selectron
production is unlikely to be detectable for any reasonable expected luminosity
at LEP2, and we consequently neglect this possibility from now on. We also
present, in Figure~\ref{nlc}, the single sneutrino production cross-section at
a Next Linear Collider operating at a centre of mass energy of $500\gev$. 

\section{Single Sneutrino Production}\label{analysis.sec}

The signal $e^+e^- \ra e \ell \tilde{\nu}$ is characterised by the electron
continuing along the beam pipe, so that the only particles visible in the
detector are the charged lepton $\ell$ and the decay products of the sneutrino.
The sneutrino can either decay {\it directly}, $\tilde{\nu} \ra e \ell$, or
{\it indirectly} via lighter neutralinos or charginos, eg. $\tilde{\nu} \ra \nu
\chi^0, \ell^-\chi^+$. In the following analysis we only consider the direct
decays and the indirect decays to neutralinos for simplicity.  The final state
depends upon both the coupling involved and the flavour of sneutrino produced.
This information is summarised in Table~\ref{finalstate}. 

\begin{table}
\centering
\begin{tabular}{|c|c|c|c|c|}
\hline
Coupling & \multicolumn{2}{c|}{Direct Decays} & \multicolumn{2}{c|}{Indirect 
Decays} \\ \cline{2-5}  
         & $\tilde{\nu}_{\mu}$ & $\tilde{\nu}_{\tau}$ & $\tilde{\nu}_{\mu}$ & 
$\tilde{\nu}_{\tau}$ \\ \hline
 122     &    $e \mu^+\mu^-$   &         -         &  $\mu \nu  \chi$ &- \\
 123     &   $e \tau^+ \tau^-$ &         -         &  $\tau \nu  \chi$& - \\
 132     &          -          &  $e \mu^+\mu^-$   &      - & $\mu\nu\chi$   \\
 133     &          -          & $e \tau^+ \tau^-$ &         - & $\tau\nu\chi$\\
 231     &  $e \tau^+ \tau^-$  &  $e \mu^+\mu^-$   &$\tau\nu\chi$&$\mu\nu\chi$\\
\hline
\end{tabular}
\caption{The final states for production of different sneutrino flavours via
different couplings for the {\it direct} decays and the {\it indirect} decays to
the lightest neutralino. The entries marked with a dash are those for which that
sneutrino flavour cannot be produced by the coupling in question.} 
\label{finalstate}
\end{table}

For a coupling $\lam_{ijk}$ the lightest neutralino can decay to the following 
final states:
\beq 
\tilde{\chi}^0_1 \ra \left\{ 
l_i^- \bar{\nu}_j l_k^+,\; l_i^+ {\nu}_j l_k^-,\; \bar{\nu}_i l_j^- l_k^+,
\; {\nu}_i l_j^+ l_k^- \right\}.
\eeq
Thus the signal for sneutrino production followed by an indirect decay to the 
lightest neutralino contains three charged leptons and two neutrinos. The latter
contributes substantial missing transverse momentum. 

To investigate the viability of searching for these signals we have written a
Monte Carlo capable of generating the different final states and included an
interface to JETSET \cite{JETSET} for the decays and to PHOTOS \cite{PHOTOS}
for final state radiation. Several simple analyses to discriminate the signal
from the dominant backgrounds were developed. PHOT02 \cite{PHOT02} was used to
generate tagged $\gamma\gamma \ra \mu^+\mu^-$ and untagged $\gamma\gamma \ra
\tau^+\tau^-$ and PYTHIA \cite{JETSET} to generate $Zee$, $We\nu$, $ZZ$ and
$W^+W^-$. Two photon processes and $Zee$ are the most important backgrounds. No
attempt was made to include the effects of detector performance, but the
acceptance of a typical LEP experiment was accounted for in the following
simple manner. The energy of a particle was assumed to be detected if the
particle had a polar angle greater than $30\mrad$ from the beam axis. Charged
particles were assumed to be tracked over the polar angle range $|\cos\theta| <
0.95$ and electrons and muons were only allowed to be identified as such if
they fell within this range. 

For the direct decays two analyses were developed to search for decays to $e\mu$
and to $e\tau$. Similarly, two analyses were developed for the best ($\lambda_{
122}$) and worst ($\lambda_{133}$) case indirect decays. These are summarised 
in Table~\ref{analysis}. 

\begin{figure}
\centering
$
\begin{array}{cc}
\epsfig{file=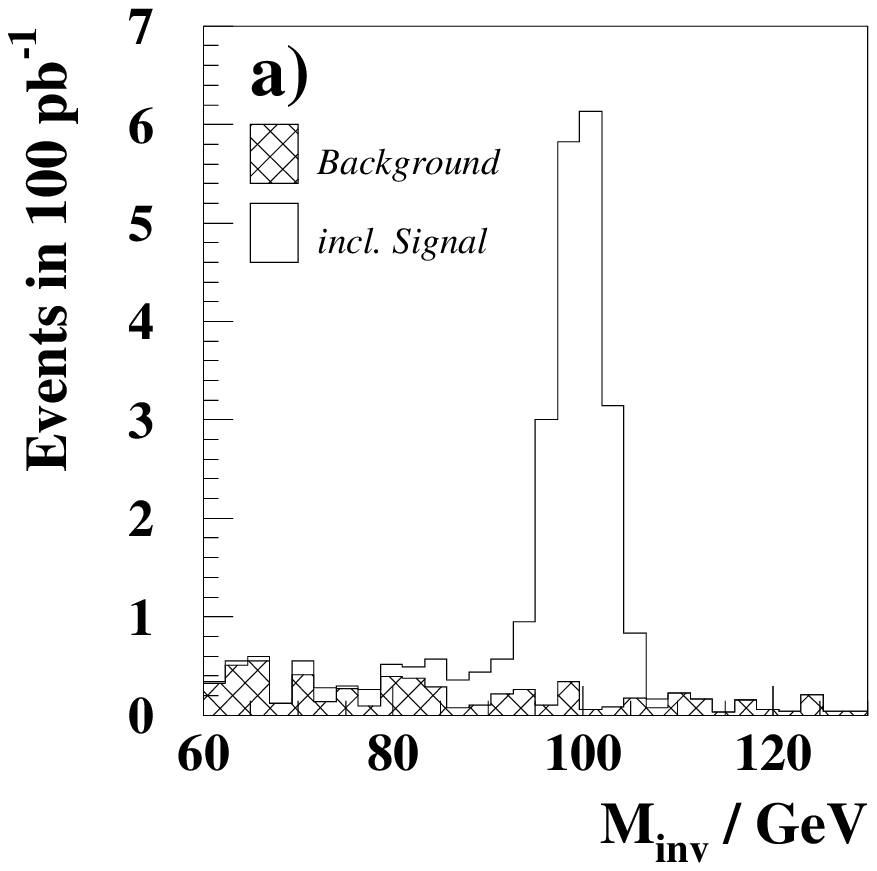,width=6.5cm} & \epsfig{file=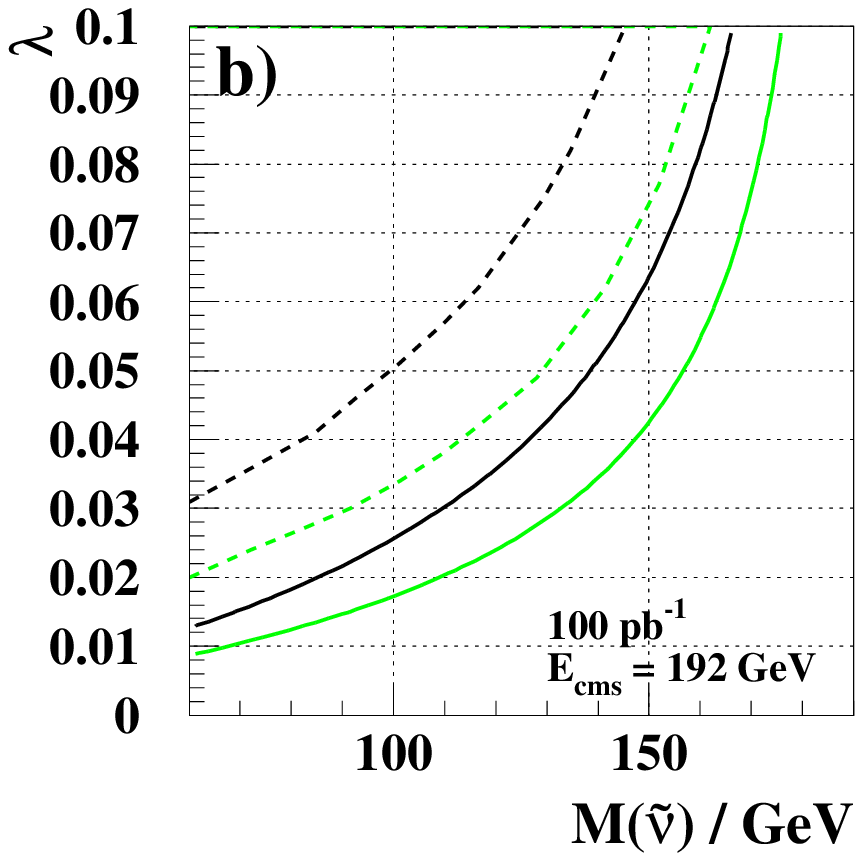,width=6.5cm} 
\end{array}
$
\caption{a) shows the distribution of the invariant mass of the
electron and muon from a sneutrino decay with $M_{\tilde\nu}=100\gev$, compared 
to the expectation from the combined backgrounds for an integrated luminosity of
$100\pbi$ and a coupling of $\lam=0.05$. In b) the discovery
contours are shown in the $(m_{{\snu}}, \lam)$ plane. The solid lines are for
the best case (indirect decays with $\lam_{122}$) and the dashed are for the
worst case (direct decays to $e\tau$). The black lines correspond to a single
experiment and the grey lines are for a combination of all four LEP
experiments.} 
\label{fig:discovery}
\end{figure}

Direct $\rpv$ sneutrino decays have a similar signature to $\gg\ra\mu\mu $ or
$\tau\tau$ with a single tagged beam electron. In the case of the signal, the
electron is produced from the decay of a massive sneutrino and is typically
emitted with a large angle, $\phi_e$, with respect to the beam axis. In
addition, the average tranverse momentum of the charged tracks, $\langle p_T
\rangle$, is larger than for a typical $\gg$ event. For the direct decays to
$e\mu$ a large visible mass is required and the invariant mass of the
sneutrino, $m_{\snu }$, can easily be constructed.  Only values of
$m_{\snu}>60\gevcc$ that are not already excluded \cite{ALEPH} are considered.
It would be possible to use the distribution of $m_{\snu}$ to improve the
sensitivity to the signal. This distribution is shown in
Figure~\ref{fig:discovery} for a sneutrino mass of $100\gev$ and a coupling of
$\lam =0.05$, assuming an invariant mass resolution of $2.5\gev$. The signal
peak is clearly visible above the background. Large missing transverse
momentum, $\not\!\! p_T$, is required for the direct decays to $e\tau$. 

For the indirect decays via a single neutralino substantial missing energy is
expected because of the presence of an energetic neutrino. This neutrino also
means that the polar angle of the missing momentum, $\theta_{m}$, is not
generally along the beam direction. The decay products of the neutralino depend
upon the choice of coupling. For a $\lam_{122}$ coupling there are three
charged leptons in the event and this can be used to reduce the background. For
a $\lam_{133}$ coupling the visible mass and total charged energy are required
to be small. The analyses, the remaining backgrounds and the efficiencies to
select a signal of a sneutrino of $100\gev$ and, for the indirect decays, a
neutralino of $50\gev$ are shown in Table~\ref{analysis}. The best performance
is achieved for the indirect decays and a coupling $\lam_{122}$; the analysis
for direct decays to $e\tau$ has the worst performance due to the larger
background. 
\begin{table}
\centering
\begin{tabular}{|l|c|c|c|c|} \hline
  & Direct $e\mu$        & Direct $e\tau$ & Indirect $\lam_{122}$&
Indirect $\lam_{133}$ \\ \hline 
& $N_{ch} = 3$ & $N_{ch} = 3$ or 5    & $N_{ch}= 3$ & $N_{ch} = 3$ or 5 \\
Cuts & $2\mu + 1e$ & $N_e = 1$ or 2 & $N_e + N_\mu = 3$ & $N_e+N_\mu \geq 1$ \\
& $\langle p_T \rangle > 15\gev$ & $\langle p_T \rangle > 15\gev$ & $\not\!\! 
p_T > 10\gev$& $\not\!\! p_T >10\gev$\\
& $|\cos\phi_e| < 0.8$ & $|\cos\phi_e| < 0.8$ & $|\cos\theta_{m}|<0.95$& 
$|\cos\theta_{m}|<0.95$      \\
& $M_{vis} > 80\gev$   & $\not\!\! p_T > 10\gev$    && $5 < M_{vis} < 60$  \\
&$M_{{\snu}} > 60\gev$  & && $E_{ch} < 70\gev$        \\ \hline 
Bkgrd. & $27\fb$           & $36\fb$    & $12\fb$ & $21\fb$ \\ \hline
Eff. & $45\%$  & $43\%$               & $49\%$   & $42\%$ \\ \hline 
\end{tabular}
\caption{Summary of the simple analyses used to derive the possible discovery 
limits. The efficiencies for a $100\gev$ sneutrino are also shown. For the case
of the indirect decays the efficiency assumes a neutralino of $50\gev$.}
\label{analysis}
\end{table}

Using these simple selection criteria and parameterising the variation of the
efficiency with sneutrino mass we can derive expected $5\sigma$ discovery
contours in the $(m_{\tilde{\nu}},\lam)$-plane. The  discovery criterion is
defined as follows: For a number of signal events, $S$, and background events,
$B$, we find $S$ such that the probability that the background fluctuates to
$S+B$ or more is less than $5.7\times10^{-5}$, i.e. $5\sigma$ from the
expectation. In Figure~\ref{fig:discovery} these contours are shown assuming
$100\pbi$ of data are collected at a centre of mass energy of $192\gev$. Much
better performance could be obtained for the case of direct decays to $e\mu$ by
including the mass distribution of the events. 
These results should be compared with the available limits upon $\rpv
$-couplings. The best current limits on $\lam_{ijk}$ in Eq.\ref{LLEsup} 
are\footnote{The bounds on $\lam_{12n}$ are at the $2\sigma$ level; all others 
are at $1\sigma$.}~\cite{bhatt} 
\begin{equation}
\begin{array}{ccccc}
\lam_{12n} < 0.05 & \lam_{131} < 0.04 & \lam_{132} < 0.04 & \lam_{133} < 0.004 
& \lam_{23n} < 0.05,
\end{array}
\end{equation}
for the right-handed slectron mass of the third index in $\lam_{ijk}$ ${\tilde 
m}_{e^c_k}= 100 GeV$. (The bounds directly depending on the sneutrino mass are
weaker.) All but the specific bound on $\lam_{133}$ scale as $(m_{{\tilde
e}^c_k}/100\gev)$. If we compare these bounds with the results presented in
Figure~\ref{fig:discovery} we see that there is a substantial discovery
potential for this new process. 

\section{Conclusion}

We have calculated the matrix elements for $e\gamma \ra {\snu}_j e_k$ and
$e\gamma \ra \slep_k \bar{\nu}_j$ via an R-parity violating coupling of type
$\lle$ and obtained the cross-sections in $\epem$ collisions. The cross-section
for single charged slepton production is too small to be investigated at LEP2.
The expected final states from single sneutrino production have been listed and
a preliminary investigation of the sensitivity to these signals made. In view
of the encouraging results derived here, a future experimental analysis to
address this possibility is very welcome. 

\section*{Acknowledgements}

We thank M. Seymour and M. Kr\"amer for very helpful discussions.

\end{document}